\documentclass[pra,amsmath,amsfonts,twocolumn,superscriptaddress,showpacs,showkeys,article]{revtex4-1}
 \usepackage{epsf,epsfig}
 \usepackage[psamsfonts]{amssymb}
\usepackage{amsthm}
\usepackage{graphicx}
\usepackage{color,soul}
\usepackage{xcolor}

 \newcommand{\ket}[1]{|#1\rangle}
 \newcommand{\bra}[1]{\langle #1|}

 \newcommand{\Id}{{\mathbb I}}
 \newcommand{\e}{{\mathrm e}}

 \newcommand{\Imagin}{{\mathrm {Im}}}
 \newcommand{\Real}{{\mathrm {Re}}}

\newcommand{\dif}{{\mathrm d}}
\newcommand{\Tr}{{\mathrm {Tr}}}

\newcommand{\etal}{\textit{et al.} }

\begin{document}
\preprint{APS/123-QED}

\title{Revealing quantum correlation  by negativity of the Wigner function}

\author{Razieh Taghiabadi}
\email{ra.taghiabadi@stu.um.ac.ir}
\affiliation{Department of Physics, Ferdowsi University of Mashhad, Mashhad, Iran}
\author{Seyed Javad Akhtarshenas}
\email{akhtarshenas@um.ac.ir}
\affiliation{Department of Physics, Ferdowsi University of Mashhad, Mashhad, Iran}
\author{Mohsen Sarbishaei}
\email{sarbishei@um.ac.ir}
\affiliation{Department of Physics, Ferdowsi University of Mashhad, Mashhad, Iran}

\begin{abstract}
We analyze  two  two-mode continuous variable separable states with the same marginal states.  We adopt the definition of classicality in the form of well-defined positive Wigner function describing the state and  find that although the states possess positive local Wigner functions, they  exhibit negative Wigner functions for the global states. Using the negativity of Wigner function as an indicator of nonclassicality, we show  that
despite these states possess different  negativities of the Wigner function, they do not reveal  this difference as phase space nonclassicalities such as negativity of the Mandel $Q$ parameter or quadrature squeezing.
We then concentrate  on quantum correlation of these states and show that quantum discord and local quantum uncertainty, as two well-defined measures of quantum correlation,  manifest the difference between negativity of the Wigner functions.
The non-Gaussianity of these states is also examined and show that the difference in behavior of their non-Gaussianity is the same as the difference between negativity of their Wigner functions.  We also investigate the influence of correlation rank criterion and find that when the states can be produced locally from classical states, the Wigner functions can not reveal their quantum correlations.

\end{abstract}


\maketitle
\section{Introduction}
In continuous variables  quantum systems, the nonclassicality of a state is closely related to the nonclassicality of the corresponding quasiprobability distribution functions in phase space \cite{CahillPR1969}.     Because of the uncertainty principle, it is not possible to define a unique well-defined distribution function, however, there exists a continuous family of quasiprobability distribution functions
\begin{equation}\label{Fs}
F^{(f)}(\alpha)=\int \frac{\dif^2 \lambda}{\pi^2} \exp{\left(\alpha \lambda^*-\alpha^* \lambda+\frac{f}{2}|\lambda|^2\right)} \Tr(D(\lambda) \rho),
\end{equation}
ranging  from the Husimi function $Q(\alpha)=F^{(-1)}(\alpha)$ to  the Glauber-Sudarshan function  $P(\alpha)=F^{(+1)}(\alpha)$
as $f$ ranges from $-1$ to $+1$. Here $D(\lambda)=\exp{\left(\lambda a^\dagger-\lambda^\ast a\right)}$ is  the displacement operator, with $a$ and $a^\dagger$ as annihilation and creation operators satisfying $[a,a^\dagger]=1$.
The Husimi $Q$ function \cite{HusimiPPMSJ1940} is defined by $Q(\alpha)=\bra{\alpha}\rho\ket{\alpha}/\pi$ and the Glauber-Sudarshan $P$ function \cite{GlauberPR1963,SudarshanPRL1963} is a diagonal representation of the state operator $\rho$ in the coherent state $\ket{\alpha}$, i.e.   $\rho=\int\dif^2\alpha P(\alpha)\ket{\alpha}\bra{\alpha}$.
The case $f=0$ corresponds to the Wigner function \cite{WignerPR1932} and  is an especially important case because integrals of the Wigner function yield
the correct quantum mechanical marginal distributions for position and momentum \cite{LeePR1995}.  Although the Husimi $Q$ function  can never become negative, it is not the case for the other ones, in the sense that there exist states for which $F^{(f)}(\rho)$ is not a regular distribution function for any $f$.   In particular,  the Wigner function $W$ is not always positive and  the Glauber-Sudarshan $P$ function can become negative or more singular than a delta function over some region of phase space. A state whose Wigner function takes on negative values over some region of phase space is nonclassical, however, the converse is not necessarily true, i.e. there exist states with positive Wigner function which show nonclassical properties.

Perhaps one of the most important application of the Wigner function is in the classification of  continuous variables quantum  states according to the classical vs nonclassical and Gaussian vs non-Gaussian paradigms \cite{HudsonRMP1974,SotoJMP1983,BrockerJMP1995,MandilaraPRA2009,GenoniPRA2013,HughesPRA2014}. For pure one- and  multi-mode states, it was proven by Hudson \cite{HudsonRMP1974} and Soto and Claverie \cite{SotoJMP1983}, respectively,  that the Gaussian states, i.e. states for which the associated Wigner function is a Gaussian function,  are the only pure states which have nonnegative Wigner functions.
Steps  toward the extension of the Hudson theorem to characterize mixed quantum states with a positive Wigner function \cite{BrockerJMP1995} have been made by finding
upper and lower bounds on the degree of non-Gaussianity of states with positive Wigner functions  \cite{MandilaraPRA2009}, and looking at criteria to detect quantum non-Gaussian states, i.e. nonclassical states that cannot be expressed as a convex mixture of Gaussian states \cite{GenoniPRA2013,HughesPRA2014}.

With the rise of quantum information science, other important signature of quantumness and the role of negativity of the  Wigner function as an indicator  have attracted attentions. It is shown that quantum non-Gaussian states with positive Wigner functions are not useful for quantum computation \cite{MariPRL2012,VeitchNJP2013}, implies that  the negativity of the Wigner function can be grasped as a resource in quantum computing and simulation.
In view of this, efforts have been made to suggest quantitative measures of nonclassicality  either as the volume of the negative part of the Wigner function \cite{KenfackJOB2004}, or as the distance  to the convex subset of positive Wigner functions \cite{MariPRL2011}.

Quantum correlation is a manifestation of nonclassicality of composite quantum systems, and a question arises as to whether the negative  Wigner function can be interpreted as a sign of the quantumness of  correlation. Two aspects of quantum correlations are quantum entanglement  \cite{EPR1935,S1935} and quantum discord \cite{ZurekPRL2001,HendersonJPA2001}. Quantum entanglement is defined within the entanglement-separability paradigm \cite{WernerPRA1989}; a bipartite state $\rho$ is entangled if it is not separable, i.e.  if  it cannot be written as a convex combination of product states as $\rho=\sum_{i}p_i\rho_i^A\otimes \rho_i^B$  where $\rho_i^{A}$ and $\rho_i^B$ are   states on $\mathcal{H}^{A}$ and $\mathcal{H}^{B}$, respectively. In the light of this,  the well-defined positive $P$ function of a two-mode state  is a sufficient condition for separability, i.e. $\rho=\int\dif^2\alpha_1\dif^2\alpha_2 P(\alpha_1,\alpha_2)\ket{\alpha_1}\bra{\alpha_1}\otimes \ket{\alpha_2}\bra{\alpha_2}$. However,  we have to mention that it's negativity does not necessarily mean entanglement because this negativity could be the result of nonclassicality of a local state or correlations beyond entanglement \cite{MarekPRA2009}.  Although the amount of entanglement is invariant under local unitary transformations, nonclassicality does not possess this property. Marek \etal \cite{MarekPRA2009} utilized this property for Gaussian states, and found local unitary operations to remove all the local manifestations of nonclassicality with the goal of reducing global nonclassicality as much as possible,  leading to the equivalence between criteria of nonclassicality and entanglement.
However, entanglement is not the only aspect of quantum correlations and  there exist  quantum correlations that can not be captured by entanglement.  In  \cite{ZurekPRL2001} Ollivier and Zurek have introduced
quantum discord as a measure of quantum correlation beyond entanglement (see also \cite{HendersonJPA2001}), and it is discovered that  it can be responsible for the quantum efficiency of deterministic quantum computation with one qubit   \cite{KnillPRL1998,DattaPRL2008}.
Quantum discord is defined from an information-theoretic perspective; a  bipartite state $\rho$ is said to have nonclassical correlation with respect to part $A$ if it cannot be distinguished locally on part $A$, i.e. if it is not possible to find orthonormal basis $\Pi_i^A=|a_i\rangle\langle a_i|$ of $\mathcal{H}^{A}$ such that $\rho$ can be written as  $\rho=\sum_ip_i\Pi_i^{A}\otimes \rho_i^{B}$. In \cite{FerraroPRL2012} it  has been shown that the set of classical-classical states, i.e. separable states that are locally distinguishable and do not possess quantum discord in any side,  and the set of states with a positive $P$ representation are almost disjoint and they are maximally inequivalent. In another word, the set of positive $P$ functions do not  often includes the set of zero-discord states, and vice versa.

An important difference between the quantum correlation within the entanglement-separability paradigm and the one described by the information-theoretic perspective is the fact that quantum entanglement can not increase under local operations and classical communication (LOCC) but quantum discord would be increased.  Moreover,  it has been pointed out that \cite{DakicPRL2010} local operations performed on a classical state can produce  a state with nonzero quantum discord. Very recently, Mani \etal  \cite{ManiPRA2015} have considered  two sets of  separable Bell-diagonal states which have different nonzero quantum correlations, although they are prepared by the same type of quantum operations acting on classically correlated states with equal classical correlations.  They have investigated this difference and found that it is related to the hidden classical correlation which is needed for preparation of these states.

In this paper we focus on the possible connection between nonclassicality in phase space  and nonclassical correlation. In particular, we concentrate  on  the negativity of the Wigner function  as an indicator of the nonclassicality in phase space and the  nonclassical correlation defined from an information-theoretic aspect, i.e. quantum discord. To this aim we introduce two two-mode separable states with nonzero quantum discord, and study their nonclassicality both from the nonclassicality in phase space  and the existence  of nonclassical correlations. Our motivation to choose these states is that: (i) Both states are classical within the  entanglement-separability  paradigm, but they are nonclassical when we look them  from  the information-theoretic perspective. (ii) The states possess the same marginal states with positive Wigner functions. (iii) The Wigner functions of these states are negative with different values of negativity. Having these states, we first look at various manifestation of phase space nonclassicalities and find that although these states possess different negativities  of the Wigner function, they do not manifest this difference as the phase  space nonclassicalities   such as negativity of the Mandel parameter or quadrature  squeezing.  We then focus our attention to the quantum correlations of these states and quantify their quantum correlations by  original quantum discord \cite{ZurekPRL2001}, local quantum uncertainty (LQU) \cite{GirolamiPRL2013}, and geometric discord \cite{DakicPRL2010}. We find that quantum correlation reveals  this difference in the sense that the state with more negativity of the Wigner function possesses more quantum correlation, measured by either  the  quantum discord \cite{ZurekPRL2001} or by LQU.  The non-Gaussianity of these states is also examined and it is shown that the difference between their non-Gaussianity is the same as the difference between negativity of their Wigner functions.

Further,   we study the role of the correlation rank criterion in revealing quantum correlation by negativity of the Wigner function. For this purpose we introduce two other separable states with  the same nonzero quantum discord. Moreover, in contrary to the previous case, the introduced states are such that the associated correlation rank of these states is 2, so that  their quantum correlations can be created  by local channels \cite{DakicPRL2010}. In this case, our calculations show that  both states and their marginals possess   the same positive Wigner functions. Our results show that the ability of Wigner function to capture the  quantum correlation  is failed when  the states can be produced locally from classical states.

Finally, we consider two classical states which  possess the same positive local Wigner functions, but their  global  Wigner functions are negative. The negativity of global Wigner function of these states are different,  however they do not exhibit squeezing and their Mandel Q parameters take the same negative values.  Looking at the states show that the states change to each other by local unitary transformation, an operation that leaves invariant classical, quantum and total correlations.  So that the
difference between negativities of these states can not be explained by aforementioned nonclassicalities. This lead us to introduce a new quantity using a local unitary operation that bring local subsystems to classical and, at the same time, reduce the amount of the negativity of the Wigner function of the  state as much as possible \cite{MarekPRA2009}. By definition such  defined quantity is invariant under local unitary operations performed on the subsystems and   associate to each state, up to a local unitary operation, a unique measure of nonclassicality.

The remainder of this paper is organized as follows. In section II we introduce two two-mode states  and study their Wigner functions. The negativity of their Wigner functions as well as their Mandel parameters and degrees of squeezing are also examined in this section.     Section III is devoted to the quantum correlations of these states. In section IV, we study  two quantum correlated states with rank 2 and study the role of  correlation rank criterion.  We conclude the paper in section V with a brief discussion.

\section{Two discordant separable states}
An important class of states of the continuous variables  quantum systems is the so-called coherent states  \cite{GlauberPR1963,SudarshanPRL1963,KlauderAP1960}.  Coherent states  result from applying the displacement operator $D(\alpha)$ to the vacuum state $\ket{0}$ of the quantized field, i.e. $\ket{\alpha}=D(\alpha)\ket{0}$  for $\alpha\in \mathbb{C}$. They  constitute the overcomplete  set of the  eigenstates of the  annihilation operator $a$, i.e. $a\ket{\alpha}=\alpha\ket{\alpha}$, and can be expressed in terms of the photon-number states $\{\ket{n}=\frac{(a^\dagger)^n}{\sqrt{n!}}\ket{0}\}_{n=0}^{\infty}$ as $\ket{\alpha}=\e^{-\frac{1}{2}|\alpha|^2}\sum_{n=0}^{\infty}\frac{\alpha^n}{\sqrt{n!}}\ket{n}$.  Moreover, coherent states  are the only states minimizing the Heisenberg  uncertainty relation, and their photon number distribution is the Poissonian statistics, i.e. $P_n(|\alpha|^2)=|\bra{n}\alpha\rangle|^2=\textrm{e}^{-|\alpha|^2}\frac{|\alpha|^{2n}}{n!}$.

Now, let $\ket{\gamma}$ and $\ket{-\gamma}$  be two coherent states of the single-mode Hilbert space $\mathcal{H}^{s}$, ($s=A,B$). Obviously, these states are not orthogonal in the sense that $\Gamma=\bra{\gamma}-\gamma\rangle=\exp{(-2\vert\gamma\vert^{2})}$. However, one can define two orthogonal states $\ket{\gamma_e}$, $\ket{\gamma_o}$, i.e. even and odd coherent states \cite{AnsariPRA1994,DodonovPhysica1974}, as
\begin{equation}
\ket{\gamma_{e,o}}=N_{e,o}(\ket{\gamma}\pm\ket{-\gamma}),
\end{equation}
where $N_{e,o}=1/{\sqrt{2(1\pm \Gamma)}}$ is the normalization constant.

\begin{figure}[t]
\centering\includegraphics[scale=0.75]{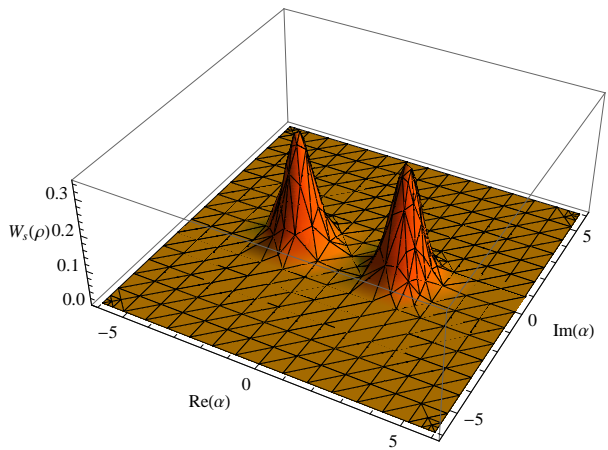}
\caption{(Color online) Local Wigner functions of the states  $\rho^{(++)}$  and $\rho^{(+-)}$,  with $|\gamma|=2$.}
\label{FigLocalWigner}
\end{figure}

With these preliminary single-mode states  in hand,  we are now in a position to define two two-mode states $\rho^{(++)}$ and  $\rho^{(+-)}$, acting on the Hilbert space $\mathcal{H}=\mathcal{H}^A\otimes \mathcal{H}^B$, as
\begin{eqnarray}\nonumber
\rho^{(++)}=&\dfrac{1}{4}\big(|\gamma\rangle\langle\gamma|\otimes|\gamma\rangle\langle\gamma|+|-\gamma\rangle\langle-\gamma| \otimes |-\gamma\rangle\langle-\gamma| \\ \label{RhoPPcv}
+& |\gamma_{e}\rangle\langle\gamma_{e}|\otimes|\gamma_{e}\rangle\langle\gamma_{e}|
+|\gamma_{o}\rangle\langle\gamma_{o}|\otimes|\gamma_{o}\rangle\langle\gamma_{o}|\big),
\end{eqnarray}
\begin{eqnarray}\nonumber
\rho^{(+-)}=&\dfrac{1}{4}\big(|\gamma\rangle\langle\gamma|\otimes|-\gamma\rangle\langle-\gamma|+|-\gamma\rangle\langle-\gamma|
\otimes|\gamma\rangle\langle\gamma| \\ \label{RhoPMcv}
+&|\gamma_{e}\rangle\langle\gamma_{e}|\otimes|\gamma_{o}\rangle\langle\gamma_{o}|
+|\gamma_{o}\rangle\langle\gamma_{o}|\otimes|\gamma_{e}\rangle\langle\gamma_{e}|\big),
\end{eqnarray}
Interestingly, both of these states have the same marginal states with respect to the modes  $A$ and $B$, i.e. defining
$\rho_{A}=\Tr_{B}[\rho]$ and $\rho_{B}=\Tr_{A}[\rho]$, we find
\begin{eqnarray}\label{RhoLS}
\rho_{A}^{(++)}&=&\rho_{B}^{(++)}=\rho_{A}^{(+-)}=\rho_{B}^{(+-)} \\ \nonumber
&=&\dfrac{1}{4}(\vert\gamma\rangle\langle\gamma\vert+\vert -\gamma\rangle\langle -\gamma\vert
+\vert\ \gamma_e\rangle\langle\gamma_e\vert+ \vert \gamma_o\rangle\langle \gamma_o\vert),
\end{eqnarray}
It turns out that the  Wigner function of the above state can be written as
\begin{eqnarray}\label{WignerSubsystems1234}
W^{\rho_A^{(++)}}&=&W^{\rho_B^{(++)}}=W^{\rho_A^{(+-)}}=W^{\rho_B^{(+-)}} \\ \nonumber
&=&\dfrac{1}{4}(W_{\gamma}(\alpha)+W_{-\gamma}(\alpha)+W_{\gamma_e}(\alpha)+W_{\gamma_o}(\alpha)),
\end{eqnarray}
where $W_{\psi}(\alpha)$ denotes the Wigner function of the pure state $\ket{\psi}$. Using Eq. \eqref{Fs} and the combination rule for the displacement operator as $D(\lambda)D(\gamma)=\textrm{e}^{(\lambda\gamma^\ast-\lambda^\ast\gamma)/2}D(\lambda+\gamma)$, we get
\begin{eqnarray}\label{WignerCS}\nonumber
W_{\gamma}(\alpha)&=&W_{-\gamma}(-\alpha)=\frac{2}{\pi}\exp{\{-2|\gamma-\alpha|^2\}}, \\ \nonumber
W_{\gamma_e}(\alpha)&=&N_e^2\left\{W_{\gamma}(\alpha)+W_{\gamma}(-\alpha)+2W_{0}(\alpha)\cos{(4\Im{{\alpha \gamma^\ast})}}\right\}, \\ \nonumber
W_{\gamma_o}(\alpha)&=&N_o^2\left\{W_{\gamma}(\alpha)+W_{\gamma}(-\alpha)-2W_{0}(\alpha)\cos{(4\Im{({\alpha \gamma^\ast})})}\right\},
\end{eqnarray}
where $W_{0}(\alpha)=W_{\gamma=0}(\alpha)$,  $\Re{\alpha\gamma^\ast}=\Real{(\alpha\gamma^\ast)}$, and $\Im{\alpha\gamma^\ast}=\Imagin{(\alpha\gamma^\ast)}$.
Clearly,  both $W_\gamma(\alpha)$ and $W_{-\gamma}(\alpha)$ are nonnegative, but  $W_{\gamma_e}(\alpha)$ and $W_{\gamma_o}(\alpha)$ may take negative values for some values of $\gamma$ and $\alpha$.  However, the sum of these latter Wigner functions become positive in the whole phase space. To see this, and for further use, let us rewrite the local Wigner function \eqref{WignerSubsystems1234} as
\begin{eqnarray}\label{WignerSubsystems}
W^{\rho_A^{(++)}}&=&W^{\rho_B^{(++)}}=W^{\rho_A^{(+-)}}=W^{\rho_B^{(+-)}} \\ \nonumber
&=&\dfrac{1}{2}(W_{q}(\alpha)+W_{c}(\alpha)),
\end{eqnarray}
where
\begin{eqnarray}\label{WignerLS12}
W_{q}(\alpha)&=&\frac{1}{2}\left(W_{\gamma}(\alpha)+W_{-\gamma}(\alpha)\right) \\ \nonumber
&=&\frac{2}{\pi}\exp{\{-2(|\alpha|^2+|\gamma|^2)\}}\cosh{(4\Re{\alpha\gamma^\ast})}, \\ \label{WignerLS34}
W_{c}(\alpha)&=&\frac{1}{2}\left(W_{\gamma_e}(\alpha)+W_{\gamma_o}(\alpha)\right) \\ \nonumber
&=&\frac{2}{\pi(1-\Gamma^2)}\exp{\{-2(|\alpha|^2+|\gamma|^2)\}} \\  \nonumber
& & \quad\quad\quad\; \times\left\{\cosh{(4\Re{\alpha\gamma^\ast})}-\cos{(4\Im{\alpha\gamma^\ast})}\right\},
\end{eqnarray}
Evidently, both $W_{q}(\alpha)$ and $W_{c}(\alpha)$ are nonnegative everywhere in phase space, so that local state \eqref{RhoLS} possesses nonnegative Wigner function, as depicted in Fig. \ref{FigLocalWigner} for $|\gamma|=2$ .

\begin{figure}[t]
\centering\includegraphics[scale=.75]{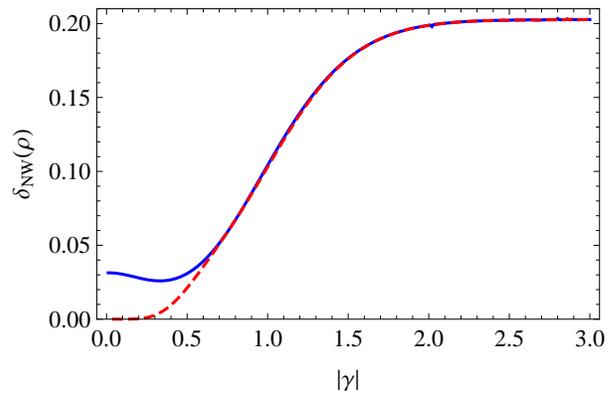}
\caption{(Color online) Negativity of the Wigner function  versus $|\gamma|$ for states $\rho^{(++)}$ (blue, solid line) and $\rho^{(+-)}$ (red, dashed line).}
\label{FigNegativity}
\end{figure}

Now in order to calculate the  Wigner function of the global states \eqref{RhoPPcv} and \eqref{RhoPMcv}, we need to use the fact that
the Wigner function of the product states  is given by the  product of the Wigner functions of each part, i.e.
$W^{\rho_1\otimes \rho_2}(\alpha_1,\alpha_2)=W^{\rho_{1}}(\alpha_{1})W^{\rho_{2}}(\alpha_{2})$. Using this we  find
\begin{eqnarray}\nonumber\label{WignerRhoPP}
W^{\rho^{(++)}}(\alpha_1,\alpha_2)&=&\dfrac{1}{4}(W_{\gamma}(\alpha_1) W_{\gamma}(\alpha_2)+W_{-\gamma}(\alpha_1) W_{-\gamma}(\alpha_2)\\ \nonumber
&+&W_{\gamma_e}(\alpha_1)W_{\gamma_e}(\alpha_2)+W_{\gamma_o}(\alpha_1)W_{\gamma_o}(\alpha_2)),
\end{eqnarray}
\begin{eqnarray}\nonumber\label{WignerRhoPM}
W^{\rho^{(+-)}}(\alpha_1,\alpha_2)&=&\dfrac{1}{4}(W_{\gamma}(\alpha_1) W_{-\gamma}(\alpha_2)+W_{-\gamma}(\alpha_1) W_{\gamma}(\alpha_2)\\ \nonumber
&+&W_{\gamma_e}(\alpha_1)W_{\gamma_o}(\alpha_2)+W_{\gamma_o}(\alpha_1)W_{\gamma_e}(\alpha_2)).
\end{eqnarray}
These Wigner functions  are not positive in the whole phase space.
Negativity of the Wigner function is a witness of nonclassicality of the state, so that measuring any departure from positivity of the Wigner function may be used as a quantitative witness of the nonclassicality of the state.  To this aim, we use the volume of the negative part of the Wigner function \cite{KenfackJOB2004} as a measure  of the nonclassicality. More precisely, the nonclassical volume is defined as a quadruple volume of the integrated negative part of the Wigner function of a two mode quantum state as \cite{KenfackJOB2004}
\begin{equation}\label{NegWigner}
\delta_{NW}(\rho)=\int\cdots\int\vert W_{\rho}(p_{1},q_{1},p_{2},q_{2})\vert \dif q_{1} \dif p_{1}\dif q_{2} \dif p_{2}-1.
\end{equation}
Using this definition, one can calculate negativity of the Wigner function for the  states above.  In Fig. \ref{FigNegativity}, we have plotted $\delta_{NW}(\rho^{(++)})$ and $\delta_{NW}(\rho^{(+-)})$ as a function of $|\gamma|$. As it is apparent from the figure, both states possess nonclassicality. Indeed, for low values of $|\gamma|$ the states have different negativities  $\delta_{NW}(\rho)$, namely $\delta_{NW}(\rho^{(+-)})<\delta_{NW}(\rho^{(++)})$,  but by increasing the value of $|\gamma|$, they reach to the same asymptotic value $\delta_{NW}(\rho^{(+-)})\approx\delta_{NW}(\rho^{(++)})\approx 0.2$. Now, we have two composite systems whose subsystems possess the same  positive Wigner functions, whereas their global  Wigner functions retain negative values. As the classicality of the subsystems is not sufficient for the nonclassicality of the overall state to be identified with entanglement \cite{MarekPRA2009}, this nonclassicality is due to various nonclassicality in phase space, such as squeezing or quantum statistics, or it may be interpreted as a signature of  other correlations, such as   classical correlation or quantum correlation beyond entanglement.

\begin{figure}
\centering\includegraphics[scale=.75]{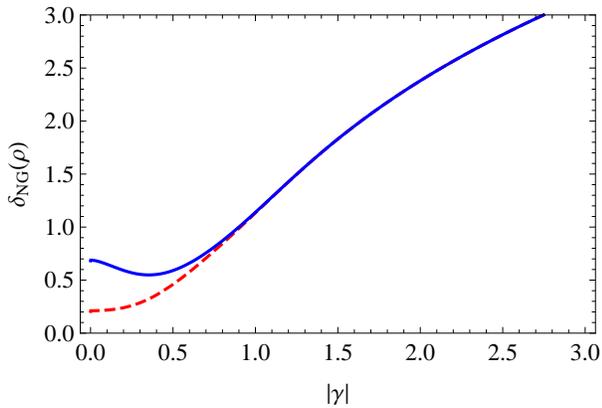}
\caption{(Color online) Quantum non-Gaussianity versus $|\gamma|$ for states $\rho^{(++)}$ (blue, solid line) and $\rho^{(+-)}$ (red, dashed line).}
\label{FigNonGaussianity}
\end{figure}

Before we proceed further to consider nonclassicality in phase space and nonclassical correlation for these states, let us investigate the nonclassical behavior of these states measuring by  non-Gaussianity. By definition, the quantum state $\rho$ is said to be Gaussian if its characteristic function or, equivalently, its Wigner function has a Gaussian form, so that    Gaussian states possess  positive Wigner functions.  Gaussian states have minimum entanglement for given second moments and non-Gaussian states are useful to improve parameter estimation \cite{GenoniPRA2009,AdessoPRA2009}.  Indeed, there exists a connection between nonclassicality and non-Gaussianity in the sense that all pure non-Gaussian states are nonclassical states, and any pure state with non-negative Wigner function is Gaussian \cite{HudsonRMP1974,LutkenhausPRA1995}.   Deviations from  Gaussian behavior are often the sign that an interesting nonclassical phenomenon occurs and several measures of non-Gaussianity were proposed \cite{GenoniPRA2007,GenoniPRA2008,GenoniPRA2010,GhiuPhysScripta2013}.  According to \cite{GenoniPRA2010}, the degree of non-Gaussianity  of a state $\rho $ is defined by
\begin{equation}
\delta_{NG}(\rho)=S(\rho\parallel\tau),
\end{equation}
where $S(\rho_1\parallel\rho_2)=\Tr{[\rho_1(\ln{\rho_1}-\ln{\rho_2})]}$ is the quantum relative entropy between states $\rho_1$ and $\rho_2$. Here $\tau$ is the reference Gaussian state with the same first and second moments of $\rho$. This property of reference state $\tau$ leads to $\Tr[\rho\ln\tau]=\Tr[\tau\ln\tau]$,  so that
\begin{equation}\label{nG}
\delta_{NG}(\rho)=S(\tau)-S(\rho),
\end{equation}
where $S(\rho)$ is the von Neumann entropy of the state $\rho$. Also $S(\tau)=h(d_{+})+h(d_{-})$ where
\begin{equation}
h(x)=(x+\dfrac{1}{2})\ln(x+\dfrac{1}{2})-(x-\dfrac{1}{2})\ln(x-\dfrac{1}{2}),
\end{equation}
and $d_{\pm}^2=\frac{1}{2}\left(\Delta(\delta)\pm\sqrt{\Delta(\delta)^{2}-4I_{4}}\right)$ are the symplectic eigenvalues of the covariance matrix $\sigma$ of the reference  Gaussian state $\tau$. Here  $\Delta(\delta)=I_{1}+I_{2}+2I_{3}$ where  $I_{1}\equiv \det(A)$, $I_{2}\equiv \det(B)$,
 $I_{3}\equiv \det(C)$, and $I_{4}\equiv \det(\sigma)$ are four local symplectic invariants of the covariance matrix
\begin{equation}
\sigma=\begin{pmatrix}A&C\\C^{\dagger}&B\end{pmatrix},
\end{equation}
where
\begin{equation}
\sigma_{jk}=\dfrac{1}{2}\left<\{R_{j},R_{k}\}\right>-\left<R_{j}\right>\left<R_{k}\right>,
\end{equation}
with $R=\lbrace q_{1},p_{1},q_{2},p_{2}\rbrace$. For the considered states $\rho^{(++)}$ and $\rho^{(+-)}$ we find
 \begin{equation}
\sigma^{\rho^{(++)}}=\begin{pmatrix} u&0&\frac{\gamma^{2}}{2}&0\\0&v&0&0\\ \frac{\gamma^{2}}{2}&0&u&0\\0&0&0&v\end{pmatrix},
\end{equation}
\begin{equation}
\sigma^{\rho^{(+-)}}=\begin{pmatrix}u&0& -\frac{\gamma^{2}}{2}&0\\0&v&0&0\\ -\frac{\gamma^{2}}{2}&0&u&0\\0&0&0&v\end{pmatrix},
\end{equation}
respectively, where $u=\frac{1}{2}\left(1+\frac{2\gamma^{2}(2-\Gamma^2)}{1-\Gamma^{2}}\right)$ and $v=\frac{1}{2}\left(1+\frac{2\gamma^2\Gamma^{2}}{1-\Gamma^{2}}\right)$. Figure \ref{FigNonGaussianity} shows the behavior of non-Gaussianity of $\rho^{(++)}$ and $\rho^{(+-)}$ in terms of $|\gamma|$. This figure shows that for low values of $|\gamma|$ we have $\delta_{NG}(\rho^{(+-)})<\delta_{NG}(\rho^{(++)})$, but by increasing the value of $|\gamma|$, they reach asymptotically to the same value.
We are now in the position to study the various nonclassicalities  of these states in phase space.

{\it Mandel parameter.---}  As it is already mentioned in section II, coherent states are the only states that their photon number distribution is the Poissonian statistics,  i.e. the variance of the photon number distribution is equal to the expected value of the photon number. It follows that although a coherent state possesses   a maximum coherence between the photon numbers,   all the photon number detections are independent. On the other hand, when  for the same mean photon number  the distribution has a larger  or smaller variance than a Poissonian distribution, the distribution is called  super-Poissonian or sub-Poissonian, respectively.  The Mandel  $Q$ parameter measures the departure of the photon statistics of the field  from the Poissonian statistics \cite{BukhariActaPPB2011} in the sense that $Q>0$ and $Q<0$ correspond to the super-Poissonian and sub-Poissonian distribution, respectively.     The super-Poissonian statistics refers to the photon bunching effect which can be described by classical optics \cite{HanburyNature1956}. On the other hand, the sub-Poissonian statistics  is  a signature of photon antibunching \cite{KimblePRL1977}, a nonclassical characteristic of light with photons more equally spaced than a coherent laser field.

\begin{figure}
\centering\includegraphics[scale=.75]{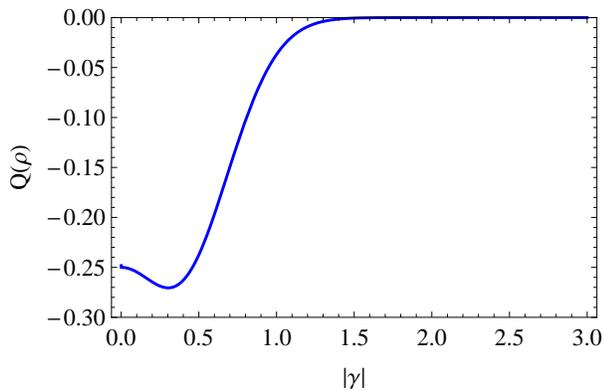}
\caption{Mandel $Q$ parameter versus $|\gamma|$ for the states $\rho^{(++)}$ and $\rho^{(+-)}$.}
\label{FigQparameter}
\end{figure}

The Mandel  $Q$ parameter is defined as the normalized variance of the photon  distribution and in each mode is defined by
\begin{equation}\label{Qparameter}
Q_{s}=\dfrac{\langle a^{\dagger^{2}}_{s}a^{2}_{s}\rangle-\langle a^{\dagger}_{s} a_{s}\rangle^{2}}{\langle a^{\dagger}_{s} a_{s}\rangle}, \quad (s=A,B).
\end{equation}
Hence, the Q parameter for both of the states $\rho^{(++)}$ and $\rho^{(+-)}$ takes the same value
\begin{equation}\label{QsRho}
Q_{s}{(\rho^{(++)})}=Q_{s}{(\rho^{(+-)})}=-|\gamma|^2 \Gamma^2\frac{2-\Gamma^2}{1-\Gamma^2},
\end{equation}
for $s=A,B$, which clearly takes negative value for all $|\gamma|$.
As a matter of fact the notion of Mandel parameter defined by Eq. \eqref{Qparameter}  for each mode of the two-mode states is a trivial extension of the definition formulated primarily \cite{MandelOL1979} for one-mode situations. Indeed, the sub-Poissonian statistics of a two-mode state may in general manifested in the linear combination of the modes different from the mode chosen for the analysis \cite{ArvindJPA1996}.
In view of this, Arvind $\etal$  \cite{ArvindJPA1996} have defined the   Mandel $Q$ parameter  for the $SU(2)$ transformed mode
\begin{eqnarray}
a(\alpha)=\alpha_{1}^{*}a_{1}+\alpha_{2}^{*}a_{2}, \quad a(\alpha)^{\dagger}=\alpha_{1}a_{1}^{\dagger}+\alpha_{2}a_{2}^{\dagger},
\end{eqnarray}
of  a two-mode state $\rho$ as
\begin{equation}\label{QparameterSU(2)}
Q(\rho;\alpha)=\dfrac{\langle a^{\dagger^{2}}(\alpha)a^{2}(\alpha)\rangle-\langle a^{\dagger}(\alpha) a(\alpha)\rangle^{2}}{\langle a^{\dagger}(\alpha) a(\alpha)\rangle}.
\end{equation}
Here  $\alpha_1$ and $\alpha_2$ are complex numbers such that $|\alpha_1|^2+|\alpha_2|^2=1$. Based on this, the signature of the nonclassical nature of $\rho$, as manifested in the photon statistics, is defined by  the $SU(2)$ invariant definition of the Mandel $Q$ parameter as \cite{ArvindJPA1996}
\begin{eqnarray}
Q(\rho)=\min_{ \alpha\in SU(2)} Q(\rho;\alpha)=Q(\rho;\overline{\alpha}),
\end{eqnarray}
where $a(\overline{\alpha})$ denotes the mode in which the sub-Poissonian statistics is manifested to the maximum degree.
Now, after some calculations, we find for both  states $\rho^{(++)}$ and $\rho^{(+-)}$ that
$\overline{\alpha}_{1}=1,\overline{\alpha}_{2}=0$ or   $\overline{\alpha}_{1}=0,\overline{\alpha}_{2}=1$, so that sub-Poissonian nature manifested by Eq. \eqref{QsRho} is the most that can be revealed by $\rho$. Figure \ref{FigQparameter} shows the $Q$ parameter of these states versus $|\gamma|$.
This figure demonstrate, however, that the considered states do exhibit same photon statistics, i.e. the Mandel parameter can not distinguish the nonclassicality of these two states exhibited by negativity of the Wigner function.

{\it Squeezing.---} Now we look at the quadrature squeezing of our considered states. Defining  the superposition quadrature operators as  \cite{LoudonMO1987}
\begin{eqnarray}
X_{1}&=&\dfrac{1}{2\sqrt{2}}(a_{1}+a_{1}^{\dagger}+a_{2}+a_{2}^{\dagger}),
\\
X_{2}&=&\dfrac{-i}{2\sqrt{2}}(a_{1}-a_{1}^{\dagger}+a_{2}-a_{2}^{\dagger}),
\end{eqnarray}
the system possesses squeezing in one of the  quadratures $1$ or $2$ if condition $(\Delta X_{1})^{2}<\dfrac{1}{4}$  or $(\Delta X_{2})^{2}<\dfrac{1}{4}$ is satisfied. Equivalently, one  can describe the presence of squeezing in quadrature $1$ or $2$ if condition
$D_{j}=4(\Delta X_{j})^{2}-1<0$ is satisfied for $j=1$ or $2$. For the states $\rho^{(++)}$ and $\rho^{(+-)}$ we find
\begin{eqnarray}
D^{\rho^{(++)}}_{1}=D^{\rho^{(+-)}}_{1}+4\gamma_x^2, \quad
D^{\rho^{(++)}}_{2}=D^{\rho^{(+-)}}_{2}+4\gamma_y^2,
\end{eqnarray}
and
\begin{eqnarray}
D^{\rho^{(+-)}}_{1}=2\frac{\gamma_x^2+\gamma_y^2\Gamma^2}{1-\Gamma^2}, \quad D^{\rho^{(+-)}}_{2}=2\frac{\gamma_y^2+\gamma_x^2\Gamma^2}{1-\Gamma^2},
\end{eqnarray}
respectively, where we have used $\gamma_x=\Real{\gamma}$ and $\gamma_y=\Imagin{\gamma}$. Clearly, the two-mode quadrature squeezing for both states $\rho^{(++)}$ and $\rho^{(+-)}$ is zero, i.e. squeezing can not distinguish these two states.

\section{Quantum  correlation}
So far we have shown that the nonclassicality properties of a two-mode photon field, such as sub-Poissonian statistics and quadrature squeezing  cannot distinguish these two states. Now we turn our attention  on the quantum correlation to investigate in detail the difference between  two states.
To this aim, we first note that states $\rho^{(++)}$ and $\rho^{(+-)}$ have the two-qubit representation by defining computational  basis   $\{\ket{e}=|\gamma_{e}\rangle,\;\ket{o}=|\gamma_{o}\rangle\}$,   for an arbitrary value of $\gamma$ and for each mode $A$ and $B$. By definition, these qubit basis depends on the value of $\gamma$, so that any change in $\gamma$ leads to the new vectors  in the infinite-dimensional Fock space of the light. For example,   in the limiting case $\gamma \rightarrow 0$, the states $\ket{e}$ and $\ket{o}$  reduce  to the  vacuum  and single-photon states $\{\ket{0},\ket{1}\}$, respectively, but on the other hand in the limit of large value of $\gamma$ the coherent states $\ket{\gamma}$ and $\ket{-\gamma}$ become  macroscopically distinguishable, so that the vectors $\ket{e}$ and $\ket{o}$ reduce to the so called Schr\"{o}dinger's cat states $\frac{1}{\sqrt{2}}\left(\ket{\gamma}\pm \ket{-\gamma}\right)$, respectively.  Now, with this definitions, in the two-qubit basis $\{\ket{ee},\ket{eo},\ket{oe},\ket{oo}\}$ we find
\begin{eqnarray}
\rho^{(++)}=\left(\begin{array}{cccc}
w_{1} & 0 & 0 & w_2\\ 0 & w_2 & w_2 & 0 \\ 0 & w_2 & w_2 & 0 \\ w_2 & 0 & 0 & w_3
\end{array}\right),
\end{eqnarray}
\begin{eqnarray}
\rho^{(+-)}=\left(\begin{array}{cccc} w^\prime_{1} & 0 & 0 & -w_2\\ 0 & w^\prime_2 & -w_2 & 0 \\ 0 & -w_2 & w^\prime_2 & 0 \\ -w_2 & 0 & 0 & w^\prime_3
\end{array}\right),
\end{eqnarray}
where we have defined
\begin{eqnarray}\label{w1w2w3}
w_{1,3}&=&\frac{1}{8}[(1\pm\Gamma)^{2}+2],    \quad w_2=\frac{1}{8}(1-\Gamma^{2}), \\ \label{wp1wp2wp3}
w^{'}_{1,3}&=&\frac{1}{8}(1\pm\Gamma)^{2}, \quad \quad \quad\;  w^\prime_2 =\frac{1}{8}(3-\Gamma^{2}).
 \end{eqnarray}
For further use, let us mention here that for a general $\gamma$ these states not only are locally but also globally unitarily inequivalent, in the sense that they possess different  spectrum of eigenvalues   given by
\begin{eqnarray}\label{SpectRhopp}
\textrm{Eig}{(\rho^{(++)})}&=&\left\{0,{1}/{4},(2+\Gamma^2)/4,(1-\Gamma^2)/4\right\}, \\ \label{SpectRhopm}
\textrm{Eig}{(\rho^{(+-)})}&=&\left\{0,{1}/{4},(1+\Gamma^2)/4,(2-\Gamma^2)/4\right\}.
\end{eqnarray}
However,  in the  limit of $\gamma\rightarrow \infty$  the two states reduce to Bell-diagonal states, i.e. states with maximally mixed marginals,  and become  equivalent up to a local unitary  transformation $i\sigma_y=\sigma_z\sigma_x$ preformed on the one of the subsystems. As a result, as we will see in the following, in the limit $\gamma\rightarrow \infty$  nonclassical futures of the above states are equivalent.
Moreover, one can easily see that for an arbitrary $\gamma$ we have  $\det{(T^{\rho^{(++)}})}=\det{(T^{\rho^{(+-)}})}=0$,  where $T$ is a $3\times 3$ matrix  defined by $T_{ij}=\Tr{(\rho \sigma_i\otimes \sigma_j)}$ with $\sigma_i$ ($i=1,2,3$) as the  Pauli matrices. This, implies that the above states  are fundamentally different from the two-qubit separable states with maximally mixed marginals, considered in  \cite{ManiPRA2015}. Indeed, the states of Ref. \cite{ManiPRA2015} belong to the two different classes with regard to the sign of $\det{(T)}$, but they becomes locally equivalent if $\det{(T)}=0$.

As these states are both separable, i.e. disentangled, we therefore pay  attention to the  quantum correlation beyond the quantum entanglement. To this aim, we use  quantum discord, local quantum uncertainty (LQU), and geometric discord  as  the possible measures of quantum correlation.

{\it Quantum discord.---}
Quantum discord of a bipartite state $\rho$, acting on the Hilbert space $\mathcal{H}=\mathcal{H}^A\otimes \mathcal{H}^B$,   is defined as the difference between two classically equivalent  but quantum mechanically different definitions
of quantum mutual information as
\begin{equation}\label{QD}
\mathcal{Q}^{(A)}(\rho)=\mathcal{I}(\rho)-\mathcal{J}^{(A)}(\rho),
\end{equation}
 where $\mathcal{I}(\rho)=S(\rho^{A})+S(\rho^{B})-S(\rho)$ is the mutual information,  measuring   the total correlation  of the bipartite state $\rho$,  and $\mathcal{J}^{(A)}(\rho)=\max_{\{\Pi_i^A\}}\mathcal{J}^{\{\Pi_i^{A}\}}(\rho)$  is the classical correlation of the state $\rho$. Here $S(\cdot)$ being the von Neumann entropy, and $\rho^{A(B)}=\mathrm{Tr}_{B(A)}(\rho)$ is the marginal state corresponding to the subsystem $A(B)$.  Also $\mathcal{J}^{\{\Pi_i^A\}}(\rho)$ denotes the mutual information of the same system after performing  the local measurement $\{\Pi_i^A\}$ on the subsystem $A$ and can be written as follows \cite{ZurekPRL2001}
\begin{equation}
\mathcal{J}^{\{\Pi_i^{A}\}}(\rho)=S(\rho^{B})-\sum_ip_iS(\rho|\Pi_i^{A}).
\end{equation}
In this equation $\{\Pi_i^{A}\}=\{|a_i\rangle\langle a_i|\}$ is the set of projection operators on the subsystem $A$, and the conditional state $\rho|\Pi_i^{A}$ is the post-measurement state, i.e.  $\rho|\Pi_i^{A}=\frac{1}{p_i}(\Pi_i^{A}\otimes\mathbb{I}^{B})\rho(\Pi_i^{A}\otimes\mathbb{I}^{B})$, where  $p_i=\mathrm{Tr}[(\Pi_i^{A}\otimes\mathbb{I}^{B})\rho(\Pi_i^{A}\otimes\mathbb{I}^{B})]$ being the probability of the $i$-th outcome of the measurement on the subsystem $A$.  We notice here that the quantum discord is in general not symmetric under the swap of the two parties, $A\longleftrightarrow B$, i.e. measuring on $A$ rather than $B$ may induce different amounts of disturbance on the generic bipartite states.

\begin{figure}
\centering\includegraphics[scale=.85]{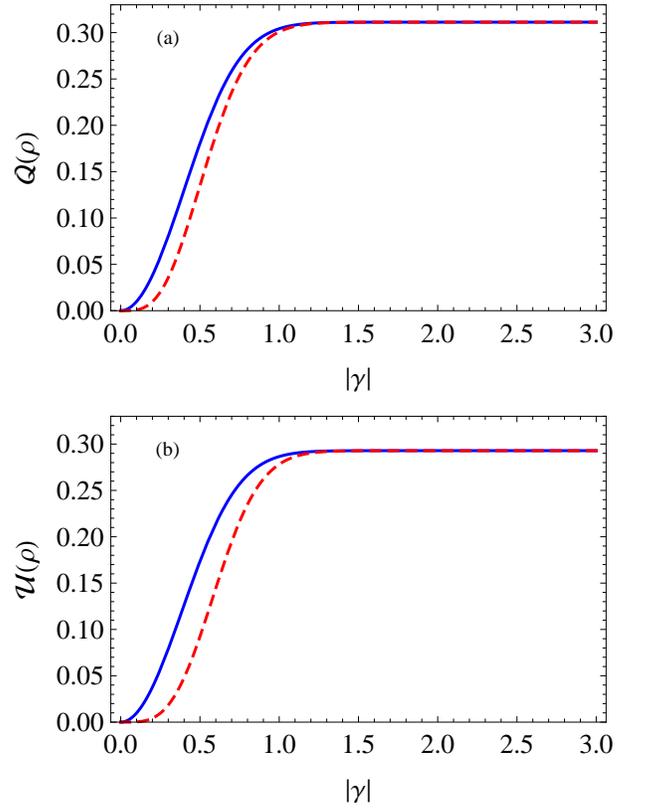}
\caption{ (a) Quantum discord and (b) LQU   versus $|\gamma|$ for the states $\rho^{(++)}$ (blue, solid line) and $\rho^{(+-)}$ (red, dashed line).}
\label{FigQD-LQU}
\end{figure}

Turning our attention to the states $\rho^{(++)}$ and $\rho^{(+-)}$, one can easily find that the optimal  measurement is  $\sigma_z$  \cite{ChenPhysicaA2011,AkhtarshenasIJTP2015}, so that
\begin{eqnarray}
\mathcal{Q}(\rho^{(++)})&=&h_4(w_1,w_2,w_3,w_2)-S(\rho^{(++)}), \\
\mathcal{Q}(\rho^{(+-)})&=&h_4(w^\prime_1,w^\prime_2,w^\prime_3,w^\prime_2)-S(\rho^{(+-)}),
\end{eqnarray}
where  $h_m(p_1,\cdots, p_m)=\sum_{i=1}^{m}-p_i \log{p_i}$ denotes  the Shannon entropy of the probabilities $\{p_1,\cdots,p_m\}$, and $w_i$ and $w^\prime_i$ are defined by Eqs. \eqref{w1w2w3} and \eqref{wp1wp2wp3}, respectively.
Figure \ref{FigQD-LQU}-a compares the quantum correlation of these states, measured by quantum discord, as a function of $|\gamma|$.   Interestingly, two states have different quantum discords.  More precisely, for low values of $|\gamma|$ ($\gamma\ne 0$) we have  $Q(\rho^{(+-)})<Q(\rho^{(++)})$,  but by increasing the value of $|\gamma|$, the discords reach to the same asymptotic value. This behavior is in agreement with the negativity of the Wigner function (Fig. \ref{FigNegativity}) as well as with the non-Gaussianity (Fig. \ref{FigNonGaussianity}), so that  quantum discord  can exhibit the difference between nonclassicality of states $\rho^{(++)}$ and $\rho^{(+-)}$.

{\it Local quantum uncertainty.---}  Local quantum uncertainty (LQU) of a bipartite state $\rho$ is defined by \cite{GirolamiPRL2013}
\begin{equation}
\mathcal{U}^{\Lambda}_{A}(\rho)\equiv \min_{K^{\Lambda}} \mathcal{I}(\rho,K^{\Lambda}),
\end{equation}
where minimum is taken over all local observables on $A$ with nondegenerate spectrum
$\Lambda$ and
\begin{equation}
\mathcal{I}(\rho,K)=-\frac{1}{2} \Tr\{[\rho^{1/2},K]^2\},
\end{equation}
is the skew information \cite{Wigner1963}. For an arbitrary state $\rho$ of a qubit-qudit system defined on $\mathbb{C}^{2}\otimes \mathbb{C}^{d}$,  $\mathcal{U}^{\Lambda}_{A}$ admits a computable closed formula as \cite{GirolamiPRL2013}
\begin{equation}\label{eqLQU}
\mathcal{U}_{A}(\rho)=1-\lambda_{\max},
\end{equation}
where $\lambda_{\max}$ denotes the maximum eigenvalue of the $3\times 3$ symmetric matrix $W$ whose matrix elements are
\begin{equation}
(W)_{ij}=\Tr\lbrace \rho^{1/2}(\sigma_{iA}\otimes \Id_{B})\rho^{1/2}(\sigma_{jA}\otimes \Id_{B})\rbrace,
\end{equation}
for  $i,j=1,2,3$. For the states  $\rho^{(++)}$ and $\rho^{(+-)}$ one can find
\begin{eqnarray}
\mathcal{U}(\rho^{(++)})&=&\frac{1-\Gamma^2}{2(1+\Gamma^2)^2}\left(2-(1-\Gamma^2)\sqrt{2+\Gamma^2}\right), \\
\mathcal{U}(\rho^{(+-)})&=&\frac{1}{1+\Gamma^2}-\frac{\sqrt{2-\Gamma^2}}{2}.
\end{eqnarray}
In Fig. \ref{FigQD-LQU}-b we have plotted the LQU  in terms of  $|\gamma|$.   Remarkably, LQU can also reveals the difference between nonclassicality of these states manifested by negativity of the Wigner function (Fig. \ref{FigNegativity}) or non-Gaussianity (Fig. \ref{FigNonGaussianity}).

{\it Geometric discord.---} The geometric discord $D_{G}$ is defined as the squared Hilbert-Schmidt distance between the bipartite state $\rho$ and the closest classical-quantum state $\chi$ \cite{DakicPRL2010}
\begin{equation}
D_{G}(\rho)=\inf_{\chi\in \Omega}\Vert\rho-\chi\Vert_{2}^{2},
\end{equation}
where $\Omega$ denotes the set of zero-discord states. For a general two-qubit state, a closed relation for the geometric discord is given in \cite{DakicPRL2010}. For the states considered in this paper, we find that both states have the same geometric discord
\begin{equation}
D_{G}(\rho^{(++)})=D_{G}(\rho^{(+-)})=\left(\frac{1-\Gamma^2}{4}\right)^2.
\end{equation}
So that, geometric discord can not manifest  the difference between the quantum correlation of these states.

The difference of negativity of the Wigner  function of states $\rho^{(++)}$ and $\rho^{(+-)}$  is only seen in quantum correlations of these states, measured by  quantum discord or LQU,  whereas all  the other nonclassicalities are equal  or do not exist, so that one can conclude that  this difference is caused by quantum correlation beyond entanglement. However, as it is clear from the figures  \ref{FigNegativity} and \ref{FigQD-LQU},  this conclusion is flawed for  $\gamma=0$  where the states  take   different nonclassicality of the Wigner function but possess the same quantum correlation, measured by quantum discord or LQU. As we will see below, this incompleteness in the connection between nonclassicality of the Wigner function and the quantum correlation can be explained by the notion of rank criterion.

\section{Correlation rank criterion}
\begin{figure}
\centering\includegraphics[scale=.85]{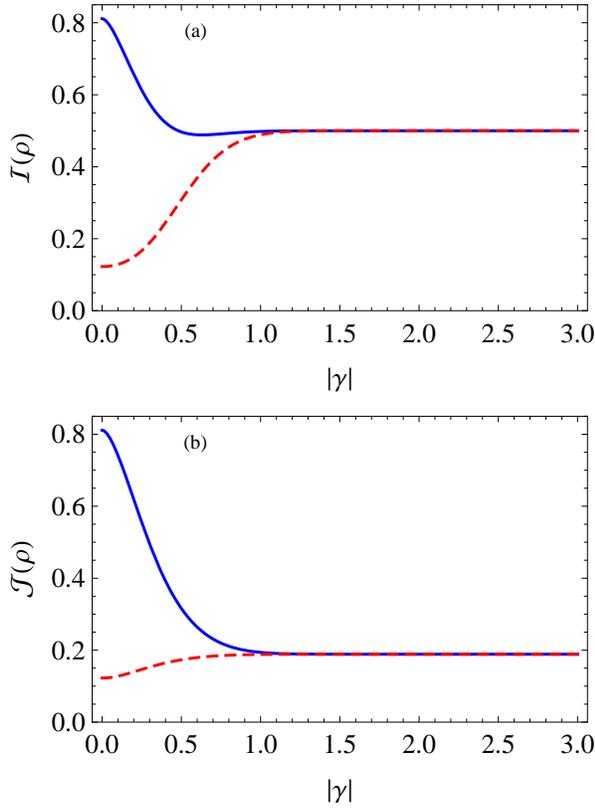}
\caption{ (a) Total  and (b) classical correlations    versus $|\gamma|$ for the states $\rho^{(++)}$ (blue, solid line) and $\rho^{(+-)}$ (red, dashed line).}
\label{FigI-J}
\end{figure}

The rank $L$ of a density operator determines the number of orthogonal operators which is needed to represent a density matrix, and  is an important discrete measure for deciding that \cite{GessnerPRA2012}:  whether  a  quantum correlated state can be produced   from  a classical state by means of   local operations, i.e. is the quantum correlation useful for quantum processing tasks. For a two-qubit classical state  we have $L\le 2$, whereas a general two-qubit quantum state can achieve all values $1\le L \le 4$. Therefore for  states with nonzero discord that can be created locally from classical states we have $L\le 2$ \cite{GessnerPRA2012}. Turning our attention to the states \eqref{RhoPPcv} and \eqref{RhoPMcv} one can see that,  when $\gamma\ne 0$,  both states  have equal rank   $L(\rho^{(++)})=L(\rho^{(+-)})=3$, i.e.  they  can not be prepared locally from classical states, however, for $\gamma=0$ the rank of both states reduces to 2, so that they can be prepared from classical states by means of local channels.
Using this fact we will argue that the disability of quantum correlation to capture the difference in nonclassicality of the Wigner functions of the states
 $\rho^{(++)}$ and $\rho^{(+-)}$ for $\gamma=0$, may be related to the locally producibility of  these states from classical states.  In the light of this, one may expect that the classical correlation can reveal this  difference, in particular for $\gamma=0$ where the states can be prepared by quantum
operations acting on the classically correlated states.
 Looking at the total and classical correlations of these states, one can find that  the states possess different nonzero
total
\begin{eqnarray}
\mathcal{I}(\rho^{(++)})&=&2S(\rho_A^{(++)})-S(\rho^{(++)}), \\
\mathcal{I}(\rho^{(+-)})&=&2S(\rho_A^{(+-)})-S(\rho^{(+-)}),
\end{eqnarray}
and classical correlations
\begin{eqnarray}
\mathcal{J}(\rho^{(++)})&=&2S(\rho_A^{(++)})-h_4(w_1,w_2,w_3,w_2), \\
\mathcal{J}(\rho^{(+-)})&=&2S(\rho_A^{(+-)})-h_4(w^\prime_1,w^\prime_2,w^\prime_3,w^\prime_2),
\end{eqnarray}
depicted in Fig.  \ref{FigI-J}, where  $S(\rho_A^{(++)})=S(\rho_A^{(+-)})=h_2(\frac{1}{2}+\frac{\Gamma}{4},\frac{1}{2}-\frac{\Gamma}{4})$.
As it is clear from Figs.  \ref{FigI-J}-a and \ref{FigI-J}-b, for small values of $|\gamma|$  both total and classical correlations of the state $\rho^{(++)}$ are larger than the corresponding quantities of the state $\rho^{(+-)}$, i.e.  $\mathcal{I}(\rho^{(++)})>\mathcal{I}(\rho^{(+-)})$ and  $\mathcal{J}(\rho^{(++)})>\mathcal{J}(\rho^{(+-)})$. More importantly when $\gamma=0$, i.e. when the states can produced locally from classically correlated states,   the classical correlation captures the difference between nonclassicality of the Wigner function, a future that is out of the scop of quantum correlation.

{\it Quantum states with correlation rank 2.---} To provide additional insight for understanding the role of rank in such states,  let us consider the following rank 2 states, obtained from the first two terms of states \eqref{RhoPPcv} and \eqref{RhoPMcv}
\begin{eqnarray} \label{SigmaPPcv}
\sigma_q^{(++)}=\frac{1}{2}\left(|\gamma\rangle\langle\gamma|\otimes|\gamma\rangle\langle\gamma|+|-\gamma\rangle\langle-\gamma| \otimes |-\gamma\rangle\langle-\gamma|\right),
\end{eqnarray}
\begin{eqnarray}\label{SigmaPMcv}
\sigma_q^{(+-)}=\frac{1}{2}\left(|\gamma\rangle\langle\gamma|\otimes|-\gamma\rangle\langle-\gamma|+|-\gamma\rangle\langle-\gamma|
\otimes|\gamma\rangle\langle\gamma|\right),
\end{eqnarray}
respectively. It is remarkable that the above states can be created locally from the classically states. To see this let us define the channel $\Phi$ as
\begin{equation}
\Phi(X)=\ket{\gamma}\bra{\gamma_e}X\ket{\gamma_e}\bra{\gamma}+\ket{-\gamma}\bra{\gamma_o}X\ket{\gamma_o}\bra{-\gamma}.
\end{equation}
Applying this channel to both modes as $\Phi\otimes \Phi$, we lead to
\begin{equation}
\sigma_q^{(++)}=(\Phi\otimes \Phi)(\sigma_c^{(++)}), \quad \sigma_q^{(+-)}=(\Phi\otimes \Phi)(\sigma_c^{(+-)})
\end{equation}
where $\sigma_c^{(++)}$ and $\sigma_c^{(+-)}$ are two classical states given  in Eqs. \eqref{SigmaCPPcv} and \eqref{SigmaCPMcv}.

For both states $\sigma_q^{(++)}$ and $\sigma_q^{(+-)}$, the Wigner function of the local states  is given by Eq. \eqref{WignerLS12}
 which clearly is positive. Also, one can easily see that both states  have  positive global Wigner functions. Moreover, the states have the same amount of quantum correlations, measured by either geometric discord
\begin{eqnarray}\nonumber
D_{G}(\sigma_q^{(++)})&=&D_{G}(\sigma_q^{(+-)}) \\
&=&\frac{1}{16}\min\left\{(1-\Gamma^2)^2,\;\Gamma^2(1+\Gamma^2)\right\},
\end{eqnarray}
or by more probing measures such as  quantum discord
\begin{eqnarray}\nonumber
\mathcal{Q}(\sigma_q^{(++)})&=&\mathcal{Q}(\sigma_q^{(+-)}) \\
&=&h_2(w_{+},w_{-})+S({\sigma_q}_s^{(++)})-S(\sigma_q^{(++)}),
\end{eqnarray}
and LQU
\begin{eqnarray}\nonumber
\mathcal{U}(\sigma_q^{(++)})&=&\mathcal{U}(\sigma_q^{(+-)}) \\
&=&\min\left\{1-\sqrt{\frac{1-\Gamma^2}{1+\Gamma^2}},\; \frac{1-\Gamma^2}{1+\Gamma^2}\right\}.
\end{eqnarray}
Here,  $w_{+}=\frac{1}{2}+\frac{1}{4}\sqrt{\Gamma^2+(1-\Gamma^2)^2}$ and $w_{-}=1-w_{+}$, and  $S(\sigma_q^{(++)})=h_2\left(\frac{1}{2}(1+\Gamma^2),\frac{1}{2}(1-\Gamma^2)\right)$ and $S({\sigma_q}_s^{(++)})=h_2\left(\frac{1}{2}(1+\Gamma),\frac{1}{2}(1-\Gamma)\right)$ are von Neumann entropies of global and local states, respectively. Moreover, in this case both states possess the same amount of classical and total correlations.
A comparison of the above results with the results we have obtained for states \eqref{RhoPPcv} and \eqref{RhoPMcv} shows that although the quantum correlations of states \eqref{RhoPPcv} and \eqref{RhoPMcv} are manifested by negativity of the Wigner function,  for  the locally producible states \eqref{SigmaPPcv} and \eqref{SigmaPMcv} the quantum correlation can not captured by Wigner function.

\begin{figure}
\centering\includegraphics[scale=.85]{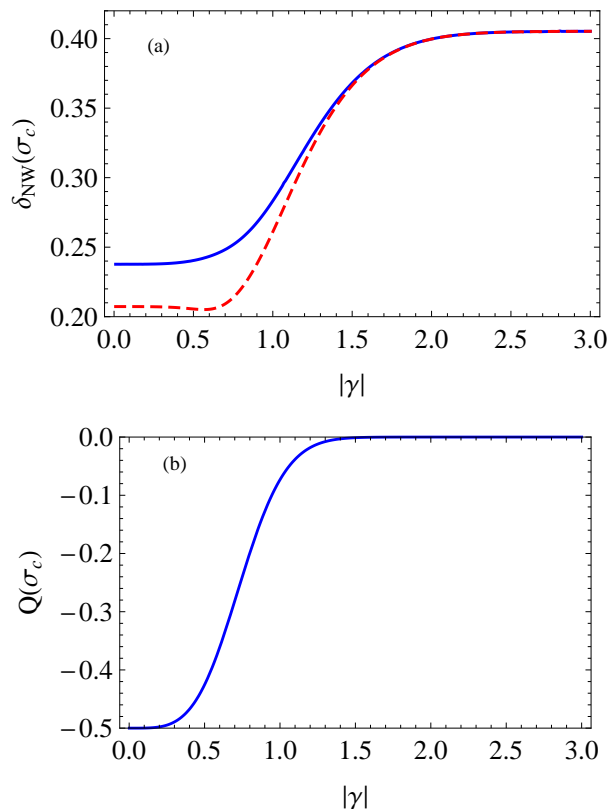}
\caption{ (a) Negativity of the Wigner function  and (b) Mendel $Q$ parameter    versus $|\gamma|$ for the states $\sigma_c^{(++)}$ (blue, solid line) and $\sigma_c^{(+-)}$ (red, dashed line). Q is the same for  both states.}
\label{FigNegQ34}
\end{figure}

{\it Classical states with correlation rank 2.---} Finally, let us consider the following two classical states, obtained from the last two terms of states \eqref{RhoPPcv} and \eqref{RhoPMcv}
\begin{eqnarray}\label{SigmaCPPcv}
\sigma_c^{(++)}=\frac{1}{2}\left(|\gamma_{e}\rangle\langle\gamma_{e}|\otimes|\gamma_{e}\rangle\langle\gamma_{e}|
+|\gamma_{o}\rangle\langle\gamma_{o}|\otimes|\gamma_{o}\rangle\langle\gamma_{o}|\right),
\end{eqnarray}
\begin{eqnarray}\label{SigmaCPMcv}
\sigma_c^{(+-)}=\frac{1}{2}\left(|\gamma_{e}\rangle\langle\gamma_{e}|\otimes|\gamma_{o}\rangle\langle\gamma_{o}|
+|\gamma_{o}\rangle\langle\gamma_{o}|\otimes|\gamma_{e}\rangle\langle\gamma_{e}|\right),
\end{eqnarray}
respectively. Local Wigner functions of these states are  positive and are given by Eq. \eqref{WignerLS34}. Moreover, both states have zero quantum correlation and the same classical and total correlations, namely $\mathcal{I}(\sigma_c^{(++)})=\mathcal{I}(\sigma_c^{(+-)})=\mathcal{J}(\sigma_c^{(++)})=\mathcal{J}(\sigma_c^{(+-)})=1$. Negativities  of these  states are different and nonzero (Fig. \ref{FigNegQ34}-a), however they do not exhibit  squeezing and their Mandel $Q$ parameters take the same negative values (Fig. \ref{FigNegQ34}-b). So that the difference between negativities of these states can not be explained by aforementioned nonclassicalities. However, a glance at these states show that they are locally equivalent, in the sense that  two states change to each other by unitary  transformation $S_x=\ket{\gamma_e}\bra{\gamma_o}+\ket{\gamma_o}\bra{\gamma_e}$  performed on the one of the subsystems, i.e. $\sigma_c^{(++)}=(S_x\otimes \mathbb{I})\sigma_c^{(+-)}(S_x\otimes \mathbb{I})$. As the amount of quantum correlation as well as classical correlation do not change by local unitary transformations,  but the nonclassicality  does not possess this property, the difference manifested in Fig \ref{FigNegQ34}-a may be related to the other nonclassicalities in phase space, such as  higher order squeezing \cite{HongPRA1985} or  the photon antibunching \cite{KimblePRL1977} which is, in general,  not equivalent to the sub-Poissonian distribution \cite{ZouPRA1990}.

{\it Minimum Negativity.---} In the  previous example we see that although the states $\sigma_c^{(++)}$ and $\sigma_c^{(+-)}$ possess different negativity of the Wigner function, this difference does  manifested neither as a phase space nonclassicality nor as a quantum correlation.  However, as we mentioned previously these states are locally equivalent. This implies  that performing any local unitary operations on these states do not influence the amount of quantum correlation, but can change the negativity of the Wigner function. Therefore we are looking  for a local unitary operation that brings local subsystems to classical and, at the same time, reduces the amount of the negativity of the Wigner function of the  state as much as possible \cite{MarekPRA2009}. By definition this associate to each state, up to a local unitary operation, a unique measure of nonclassicality. Accordingly, we define the following quantity as the minimum negativity of the state
\begin{equation}
\delta_{NW}^{\min}(\rho)=\min_{U_A\otimes U_B}\delta_{NW}((U_A\otimes U_B)\rho (U_A\otimes U_B)^\dagger).
\end{equation}
Here  minimum is taken over all local unitary operations that bring the subsystems to classical, i.e. the subsystems attain positive Wigner functions.
Using the above definition, it is clear that for the  locally equivalent states $\sigma_c^{(++)}$ and $\sigma_c^{(+-)}$ we have
$\delta_{NW}^{\min}(\sigma_c^{(++)})=\delta_{NW}^{\min}(\sigma_c^{(+-)})$.

\section{conclusion}
 We have considered  two locally equivalent, but globally different, two-mode separable states  $\rho^{(++)}$ , $\rho^{(+-)}$  and  analyzed the nonclassicality of theses states. It is shown that in spite of the posivity of the Wigner functions of  the subsystems and equality of  the global nonclassicality futures   such as  Mandel $Q$ parameter and quadrature squeezing, the negativity of the global Wigner functions for these states are different. More investigation have reveled   that quantum correlation is responsible for the difference manifested by the negativity of the Wigner function. Indeed, it is shown that  negativity of the Wigner function, non-Gaussianity, quantum discord, and LQU of the state $\rho^{(++)}$ are larger than the corresponding quantities of the state  $\rho^{(+-)}$, but the geometric discord of both states are the same.

 We have also investigated the role of rank criterion and found that when the quantum correlated states can be produced locally from classically correlated states, classical correlation, instead of quantum correlation,  can be responsible for the difference between negativity of the Wigner function of two states.
Moreover, we have also introduced two another pairs of  states, denoting by $\{\sigma_q^{(++)}, \sigma_q^{(+-)}\}$  and $\{\sigma_c^{(++)}, \sigma_c^{(+-)}\}$, with the goal that the first pair has nonzero quantum correlation but their quantum correlation can be produced by local channels, and the second pair exhibits no quantum correlation at all. More precisely, we have introduced a local channel $\Phi\otimes \Phi$ in such a way that  we can prepare the former pair from the second one.  We found that  for the first pair, the Wigner functions of  both states are positive and  the states possess the same amount of  quantum correlations measured by quantum discord, LQU, and geometric discord. On the other hand for the second pair, which contains two locally unitarily equivalent states with no quantum correlation, the Wigner functions possess different negativity, although they possess the same amount of negative  Mandel $Q$ parameter and no evidence for squeezing. In order to overcome this issue we have defined minimum negativity of a state which is invariant under local unitary operation. Further study on this measure is under considerations.

\end{document}